\documentclass[slac_one]{revtex4}
\usepackage{graphicx}
\usepackage{fancyhdr}
\begin{document}

\title{Electroweak Theory for the Tevatron, LHC, and ILC}
\author{J. Erler}
\affiliation{Departamento de F\'isica Te\'orica,
Instituto de F\'isica,
Universidad Aut\'onoma de M\'exico,
M\'exico D.F. 04510,
M\'exico}

\begin{abstract}
Future high precision electroweak measurements require understanding of
Standard Model expectations to multi-loop accuracy, both, for the prediction
of production cross-sections of signal and background, as well as for
pseudo-observables. I review recent results from precision calculations and
summarize projections and implications for the Tevatron, the LHC, CEBAF, and
the ILC.
\end{abstract}

\maketitle

\thispagestyle{fancy}


\section{INTRODUCTION}
Electroweak (EW) theory enters collider physics in three distinct places. Precision calculations at high loop orders (reviewed in Section~\ref{pseudo-observables}) are required to compute so-called pseudo-observables, such as the mass and width of the W boson or the weak mixing angle, so as to permit a meaningful comparison with their values extracted from experiment. In turn, this extraction demands another class of computations that are used (via Monte Carlo generators or explicitly applied corrections) by the experimentalists to interpret their data (see Section~\ref{predictions}).

Finally, all relevant experimental and theoretical information needs to be gathered and analyzed simultaneously.  This necessitates a careful consideration of experimental correlations; the identification of common theoretical uncertainties; a globally defined set of input and fit parameters; and the conversion of the various theoretical results to a self-consistent framework of conventions and renormalization schemes. Furthermore, constraints on new particles arise both, directly from searches and indirectly from very high precision measurements where they may leave footprints, so appropriate combination procedures must be applied. Section~\ref{combination} exemplifies this using current constraints on the Standard Model (SM) Higgs boson. Section~\ref{precision} summarizes the prospects for future precision measurements in the Higgs sector and of key EW parameters at existing and proposed accelerators.

\section{\label{pseudo-observables} ELECTROWEAK LOOP CORRECTIONS}
The four heavy weights of the SM, the Higgs, $W$, and $Z$ bosons, and the top quark, satisfy, respectively, the simple tree level relations, $M_H = \lambda v$, $M_W = g v/2$, $M_Z = {\sqrt{g^2 + {g^\prime}^2}}/2 v$, and $M_t = y_t v$. The vacuum expectation value, $v = [\sqrt{2} G_F]^{-1/2} = 246.2209 \pm 0.0005$~GeV, of the CP-even (real) Higgs field component, $H$, is determined using the latest experimental results on the $\mu$ lifetime~\cite{Chitwood:2007pa,Barczyk:2007hp} and the two-loop calculation~\cite{vanRitbergen:1998yd} to $\mu$ decay. $\lambda$ is defined relative to the Higgs potential (in unitary gauge after spontaneous symmetry breaking),
\begin{equation}
V_H = - {M_H^4\over 8\lambda^2} + M_H^2 {H^2\over 2} +
3\lambda M_H {H^3\over 3 !} + 3 \lambda^2 {H^4\over 4 !},
\end{equation}
while $g$ and $g^\prime$ are the gauge couplings and $y_t$ is the top quark Yukawa coupling (in an appropriate normalization).  These relations are modified by EW radiative corrections. E.g., the parameter $\Delta\hat\rho$ may be defined by~\cite{Degrassi:1990tu},
\begin{equation}
\Delta\hat\rho \equiv {\cos^2\theta_W \over \cos^2\hat\theta_W} - 1 \sim {3\alpha \over 16\pi\sin^2\hat\theta_W}{M_t^2\over M_W^2},
\end{equation}
showing its leading quadratic $M_t$ dependence~\cite{Veltman:1977kh}, where $\cos\theta_W \equiv {M_W\over M_Z}$, while $\cos^2\hat\theta_W = {g^2\over g^2 + {g^\prime}^2}$ may be any coupling constant based definition. In the following we use the one based on $\overline{\rm MS}$ scheme renormalization with some logarithmic $M_t$ dependence removed, so that the relation to the effective $Z$ pole mixing angle is almost a constant shift~\cite{Gambino:1993dd},
\begin{equation}
\Delta\hat\kappa_\ell \equiv  {\sin^2\theta_\ell^{\rm eff.}\over \sin^2\hat\theta_W} - 1 \sim 0.00125.
\end{equation}
The remaining parameter~\cite{Sirlin:1989uf} has a dominantly logarithmic $M_t$ dependence and gets a contribution from the renormalization group evolution (RGE) of the QED coupling to the $Z$ scale (injecting an uncertainty from quark loops),
\begin{equation}
\Delta\hat{r}_W \equiv 1 - {\pi\alpha\over \sqrt{2}G_F M_W^2 \sin^2\hat\theta_W} \sim {\alpha\over 4\pi\sin^2\hat\theta_W} \ln {M_t^2\over M_W^2} + \Delta\hat\alpha(M_Z).
\end{equation}

Except for $\Delta\hat\kappa_\ell$, the full two-loop contributions~\cite{Freitas:2000gg,Awramik:2002wn,Onishchenko:2002ve,Awramik:2004ge,Hollik:2005ns} to these parameters were completed recently. These calculations became necessary after the leading first two terms in an expansion in $M_t^{-2}$, those of
${\cal O}(M_t^4/M_Z^4)$~\cite{Barbieri:1992nz,Fleischer:1993ub} and
${\cal O}(M_t^2/M_Z^2)$~\cite{Degrassi:1996mg}, showed very poor convergence.
The leading three-loop terms of
${\cal O} (\alpha^3 M_t^6)$~\cite{vanderBij:2000cg,Faisst:2003px} and
${\cal O} (\alpha^3 M_H^4)$~\cite{Boughezal:2004ef} have also been found. The level of precision anticipated for the LHC and the ILC also requires the knowledge of mixed QCD-EW effects, especially if enhanced.  The leading contributions of this type are of
${\cal O} (\alpha\alpha_s M_t^2)$~\cite{Djouadi:1987gn,Kniehl:1988ie} and ${\cal O}(\alpha\alpha_s)$~\cite{Halzen:1990je,Fanchiotti:1992tu,
Djouadi:1993ss}. At the three-loop level, all terms of ${\cal O} (\alpha\alpha_s^2 M_t^2)$ are known, including both singlet (purely gluonic)~\cite{Anselm:1993uq} and non-singlet~\cite{Chetyrkin:1995ix,Avdeev:1994db} terms, and for the latter case also the next two terms in the expansion in $M_t^{-2}$~\cite{Chetyrkin:1995js}. Even the four-loop terms of ${\cal O} (\alpha\alpha_s^3 M_t^2)$ have been calculated, again both of singlet~\cite{Schroder:2005db} and non-singlet~\cite{Chetyrkin:2006bj,Boughezal:2006xk} types. Finally, at the level of two EW loops, the ${\cal O} (\alpha^2\alpha_s M_t^4)$ result were found in~\cite{vanderBij:2000cg}. These results ensure solid SM predictions for pseudo-observables at the LHC, and for almost most purposes even at the ILC.

\section{\label{combination} PRECISION MEASUREMENTS AND HIGGS SEARCHES}
The pre-LHC era, driven largely by the very high precision $Z$ pole programs at LEP and the SLC, witnessed a new level of direct cooperation between experimental and theoretical high energy physics. Not only provided theorists and experimentalists each other with guidance, but precision calculations entered directly into measurement processes, in many cases in each step from the planning to final analysis phases.  Conversely, experimental information was not only used but also specifically produced to pitch in wherever theory was unable to provide solid answers as showcased by the hadronic contributions to the muon $g-2$, $\Delta\hat\alpha(M_Z)$, or the RGE of $\sin^2\hat\theta_W$.  As a by-product of this mutual influence, much more time and effort is now being spent to estimate theoretical uncertainties, a task that is difficult and at the same time indispensable given that these enter straight into the quoted experimental results.

As for the future, the further increase in precision will also require to account for theoretical correlations that can occur.  This is --- once individual contributions to theory uncertainties have been obtained and agreed upon --- in fact a more straightforward exercise, but one that so far is often neglected as it can be tedious and time consuming.

\begin{figure}[h]
\includegraphics[height=62mm]{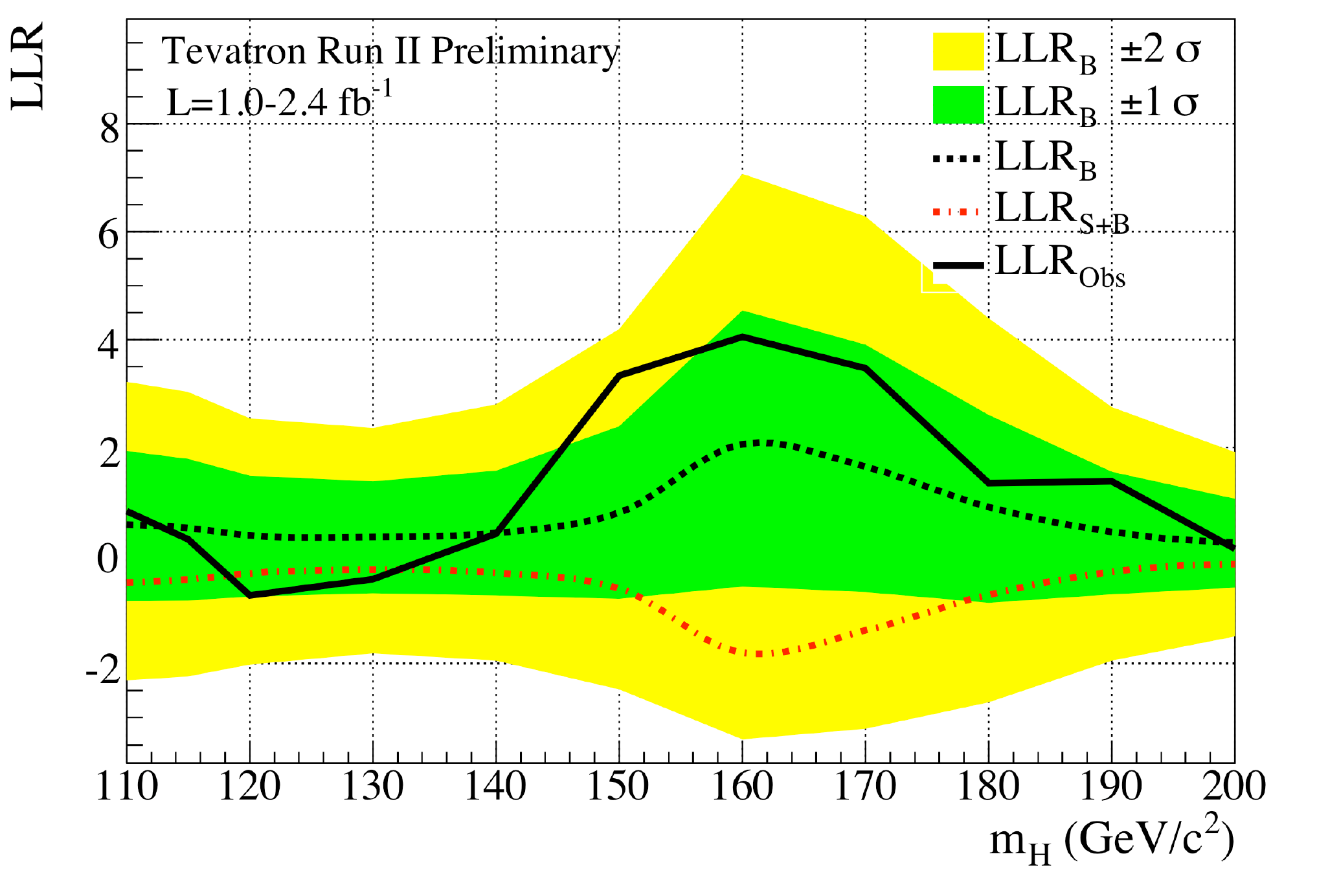}\includegraphics[height=62.1mm]{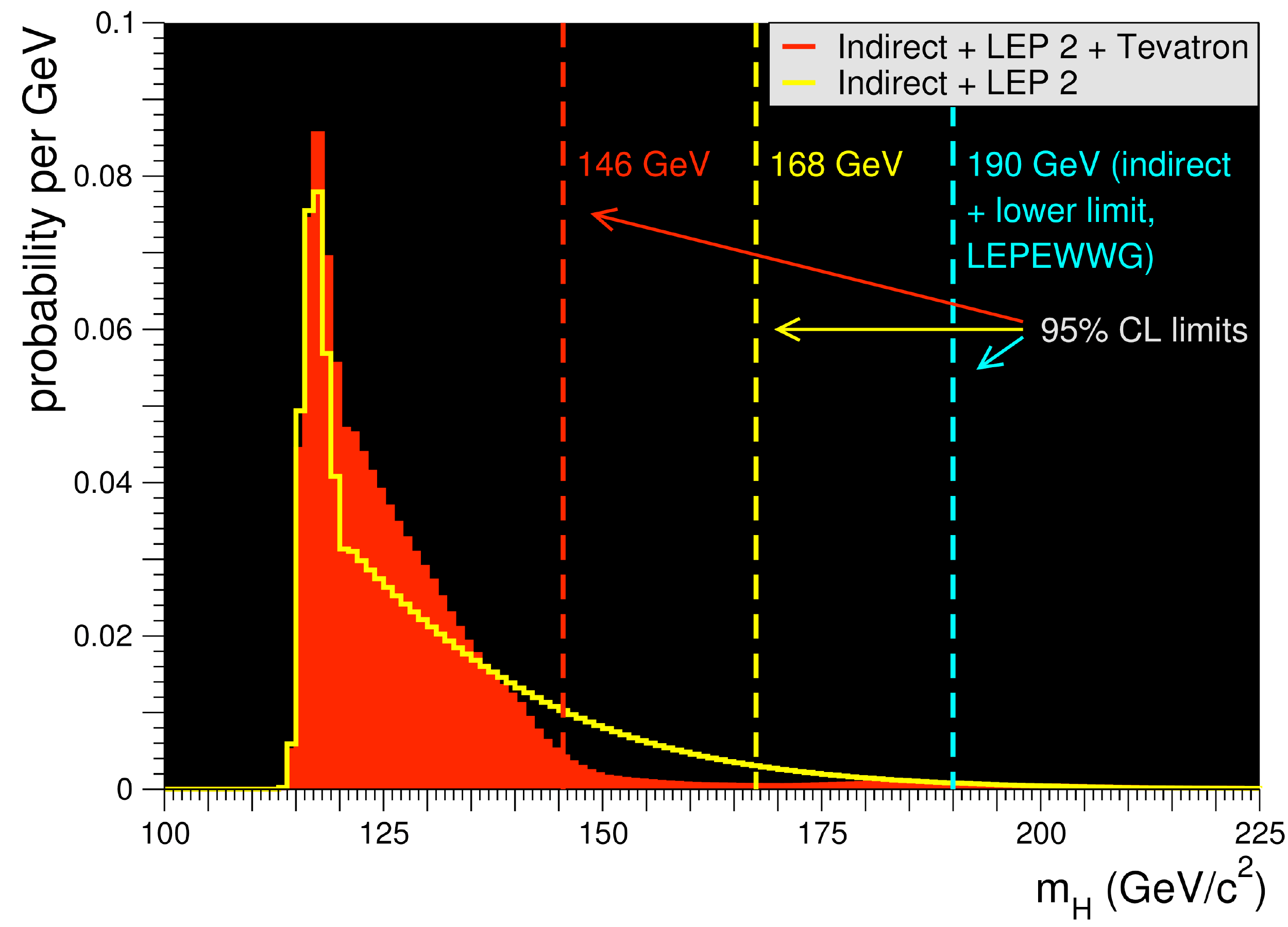}
\caption{Left: log-likelihood ratio (LLR) of probabilities conditioned on the Higgs over the background-only hypotheses as a function of $M_H$ from the Tevatron~\cite{TEVNPH:2008ds} (LEP searches~\cite{Barate:2003sz} are also included). The observed LLR (solid line) can be added to the minimum $\chi^2$-function obtained from the precision data, resulting in the probability distribution shown on the right.}
\label{tevatron-higgs}
\end{figure}

The simultaneous incorporation of information from multiple sources will also intensify. New particles will be looked for in both real production and loops. Up to now these two sectors are analyzed independently or rough (lower) bounds on new particle masses are imposed when interpreting pseudo-observables in terms of them. However, at variance with lepton colliders, the search results of the LHC and already at the Tevatron Run II cannot be approximated by sharp edged exclusion limits even in cases where no excess of events is observed, but will generally take a functional form in a multi-dimensional parameter space. An example is given in Figure~\ref{tevatron-higgs} for the SM Higgs boson.

\section{\label{precision} FUTURE PRECISION MEASUREMENTS} 
Kinematic reconstruction entirely dominates the current average of $M_W = 80.399 \pm 0.025$~GeV from LEP~2~\cite{Alcaraz:2007ri} and the Tevatron~\cite{TEVEWWG:2008ut}.  The final Tevatron combination of the $e$ and $\mu$ decay modes may yield a precision of $\pm 14$~MeV, while the LHC (after high luminosity running and the collection of about 400~fb$^{-1}$) could achieve $\pm 6$~(7)~MeV for the $e$ ($\mu$) final state alone. An uncertainty of $\pm 7$~MeV  would also be achievable with a $W^\pm$ threshold scan at an ILC.

$\sin^2\hat\theta_W$ has been measured very precisely at LEP and the SLC~\cite{LEPEWWG:2005ema}, but the long-standing $3\sigma$ conflict between the most precise values, namely from the left-right polarization asymmetry (SLD) and the forward-backward (FB) asymmetry into $b$-quarks (LEP), has never been resolved. Future determinations of $\sin^2\hat\theta_W$ of similar precision are therefore badly needed and may be possible to obtain by improving the existing measurements of the leptonic FB asymmetries at the Tevatron~\cite{Acosta:2004wq,Abazov:2008xq} and polarized M\o ller scattering at SLAC~\cite{Anthony:2005pm}, the latter being discussed as an interesting opportunity for CEBAF (JLab). The M\o ller asymmetry would also be the basis for the best determination ($\pm 7\times 10^{-5}$) at the ILC, unless the GigaZ option were realized which would yield the ultimate precision of $\pm 1.3\times 10^{-5}$. The LHC at high luminosity could also contribute a very high precision measurement comparable to the current world average provided large rapidity coverage (up to $\eta = 4.9$ for jets and missing transverse energy) can be achieved. 

For an unambiguous interpretation of these measurements one needs and expects improved determinations of $M_t$ and $\Delta\alpha(M_Z)$, calling for further experimental and theoretical efforts in both cases. In the optimistic scenario with an ILC $t\bar{t}$ threshold scan ($\delta M_t \approx \pm 50$~MeV), a GigaZ, a reduction in the error of $\Delta\alpha(M_Z)$ by a factor of three, and a high precision constraint on $\alpha_s$ (both from EW and strong interaction processes), one can determine $M_H$ from loop effects to within 4\%. This is to be compared to a $ZH$ threshold scan with a target of $\delta M_H \approx \pm 40$~MeV. Furthermore, one can compare these results with direct determinations of the Higgs self-coupling, $\lambda$, although this would require a luminosity upgrade of the LHC: 3~ab$^{-1}$ of data could determine $\lambda$ to 20\% (70\%) for $150~{\rm GeV} < M_H < 200$~GeV ($M_H < 140$~GeV). Thus, for a light Higgs one would like an ILC for which a 20\% measurement would be feasible even for $M_H = 120$~GeV. Finally, the LHC could constrain heavy fermion Yukawa and Higgs gauge couplings in the range of 10 to 30\% (even before an upgrade) while the ILC would be very precise here (albeit less so for $Hc\bar{c}$).

\section{\label{predictions} USES OF ELECTROWEAK PHYSICS IN THE FUTURE}
More recently EW physics has been moving towards becoming a tool (and background) for other objectives. E.g., one can measure the $W$ charge asymmetry and $Z$ rapidity distributions at hadron colliders to obtain information about parton distribution functions (PDFs). Or one can compute $W$ and $Z$ production cross-sections in tandem with PDFs to determine beam luminosities and detector efficiencies. 

By identifying a high $\ell^+ \ell^-$ invariant mass peak, 100~fb$^{-1}$ of LHC data could quite easily discover an extra $Z^\prime$ boson not exceeding 4 to 5~TeV in mass~\cite{Dittmar:2003ir}, depending on the underlying $U(1)^\prime$ symmetry.  To diagnose the $Z^\prime$ one can consult leptonic FB asymmetries and experiments at low energy or low momentum transfer (the $Z$ pole is rather insensitive to new physics not affecting the $Z$ couplings). Likewise, a high $\nu \ell$ transverse mass peak may reveal a $W^\prime$.

All this requires high precision predictions for single gauge boson production~\cite{Wackeroth:2006pj,Baur:2007ub,Gerber:2007xk} for which the next-to-next-to-leading order fully differential cross-section with leading-logarithmic soft gluon re-summation for transverse momenta has been completed. ${\cal O}(\alpha)$ EW corrections and final state single and mulitple $\gamma$ radiation shift the extracted $M_W$ by 10, $-168$ and 10~MeV, respectively. Open issues include higher orders in the large EW Sudakov-like logarithms, $\ln s/M_W^2$, ${\cal O}(\alpha\alpha_s)$ corrections, non-perturbative QCD contributions, as well as small $x$ and heavy quark mass effects.

\begin{acknowledgments}
This work was supported by UNAM as DGAPA-PAPIIT project IN115207.
\end{acknowledgments}

\end{document}